\documentclass[twocolumn,english,aps,prd,longbibliography,superscriptaddress]{revtex4-1}
\usepackage[T1]{fontenc}
\usepackage[latin9]{inputenc}
\usepackage{color}
\usepackage{bm}
\usepackage{xcolor}
\usepackage{amsmath}
\usepackage{amssymb}
\usepackage{graphicx}
\usepackage{natbib}
\usepackage{braket}
\usepackage{psfrag}
\usepackage{tikz}
\usepackage[normalem]{ulem}
\usepackage{qcircuit}

\usepackage[colorlinks=true,urlcolor=blue,citecolor=blue,linkcolor=blue]{hyperref}

\begin{document}


\title{Credit Risk Analysis using Quantum Computers}

\author{Daniel J.~Egger}%
\affiliation{IBM Research -- Zurich}

\author{Ricardo Garc\'ia Guti\'errez}%
\affiliation{IBM Spain}

\author{Jordi Cahu\'e Mestre}%
\affiliation{IBM Spain}


\author{Stefan Woerner}
\email{wor@zurich.ibm.com}
\affiliation{IBM Research -- Zurich}


\date{\today}

\begin{abstract}
We present and analyze a quantum algorithm to estimate credit risk more efficiently than Monte Carlo simulations can do on classical computers. 
More precisely, we estimate the economic capital requirement, i.e.~the difference between the Value at Risk and the expected value of a given loss distribution.
The economic capital requirement is an important risk metric because it summarizes the amount of capital required to remain solvent at a given confidence level.
We implement this problem for a realistic loss distribution and analyze its scaling to a realistic problem size.
In particular, we provide estimates of the total number of required qubits, the expected circuit depth, and how this translates into an expected runtime under reasonable assumptions on future fault-tolerant quantum hardware.
\end{abstract}

\maketitle

\section{\label{sec:introduction} Introduction}

Economic Capital, a key tool of risk management, is computed by financial service firms to determine the amount of risk capital that they require to remain solvant in the face of adverse yet realistic conditions \cite{Porteous2003}.
Financial service firms are exposed to many forms of risk \cite{Porteous2002} such as credit risk which is the risk of a monetary loss resulting from a counterparty failing to meet a financial obligation \cite{BIS2000, Bouteille2013}. 
For instance, a payment may not be made in due time or at all. 
Risk metrics such as Value at Risk and the Economic Capital Requirement (ECR) are often calculated for many different scenarios.
Monte Carlo (MC) simulations are thus the method of choice for this task.

In a MC simulation a parameter is estimated by building a distribution obtained by taking $M$ samples from the model input distributions.
The error on the resulting estimation scales as $\mathcal{O}(1/\sqrt{M})$ \cite{Glasserman2003}. Evaluating credit risk with MC is a rare-event simulation problem which requires many samples thereby making MC computationally costly \cite{Glasserman2005}. Importance sampling reduces the computational cost by lowering the constants but does not change the asymptotic rate of convergence.

Quantum computers process information using the laws of quantum mechanics \cite{Nielsen2010}. This opens up novel ways of addressing various computational tasks.
Problems that may benefit from quantum computing include quantum chemistry calculations \cite{Moll2018, Kandala2018}, machine learning \cite{Havlicek2019}, and finance \cite{Woerner2019, Rebentrost2018, Martin2019, Orus2019}.
Recently, it has been shown how the Quantum Amplitude Estimation (QAE) algorithm can be used to analyze financial risk measures \cite{Woerner2019} or to price financial derivatives \cite{Stamatopoulos2019} with a quadratic speedup. 

In Section \ref{sec:credit_risk_analysis}, we formally define the economic capital requirement as well as the two different uncertainty models considered.
In Section \ref{sec:quantum_algorithm}, we build on previous work \cite{Woerner2019} and discuss how to implement the quantum algorithms on a gate based quantum computer.
In Section \ref{sec:results}, we show simulation results for small instances of the considered models.
Section \ref{sec:scaling} analyzes the scaling of the algorithm for problems of realistic size as well as the resulting quantum advantage.

\section{\label{sec:credit_risk_analysis} Credit Risk Analysis}

ECR summarizes in a single figure the amount of capital (or own funds) required to remain solvent at a given confidence level (usually linked to the risk appetite or target solvency rating) and a time horizon (usually one year).
It is a complementary metric to the regulatory capital requirements that refers to the amount of own funds required following regulatory criteria and rules \cite{BaselIIIa}.
In this paper, we consider only the ECR related to default risk, which is the loss that occurs when an obligor does not fulfill the repayment of a loan.
The main components of an ECR model for a portfolio of assets are the single-asset default probabilities, the loss given default, and the correlation among the single-asset default events.
In the following, we first introduce a general form of the credit risk analysis problem considered in this manuscript and then define concrete models in detail.

For a portfolio of $K$ assets the multivariate random variable $(L_1, ..., L_K) \in \mathbb{R}_{\geq 0}^K$ denotes each possible loss associated to each asset.
The expected value of the total loss $\mathcal{L} = \sum_{k=1}^K L_k$ is $\mathbb{E}[\mathcal{L}] = \sum_{k=1}^{K} \mathbb{E}[L_k]$.
The Value at Risk (VaR) for a given confidence level $\alpha \in [0, 1]$ is defined as the smallest total loss that still has a probability greater than or equal to $\alpha$, i.e.,
\begin{eqnarray}
\text{VaR}_{\alpha}[\mathcal{L}] &=& \inf_{x \geq 0} \left\{ x \mid \mathbb{P}[\mathcal{L} \leq x] \geq \alpha \right\}.
\end{eqnarray}
The ECR at confidence level $\alpha$ is thus defined as
\begin{eqnarray}
\text{ECR}_{\alpha}[\mathcal{L}] &=& \text{VaR}_{\alpha}[\mathcal{L}] - \mathbb{E}[\mathcal{L}].
\end{eqnarray}
Common values of $\alpha$ for ECR found in the finance industry are around $99.9\%$.

In a first model, we assume that all losses are independent and can be expressed as $L_k = \lambda_k X_k$ where $\lambda_k > 0$ is the loss given default (LGD) and $X_k \in \{0, 1\}$ is a corresponding Bernoulli random variable.
The probability that $X_k=1$, i.e., a loss for asset $k$, is $p_k$.
The expected loss of the portfolio $\mathbb{E}[\mathcal{L}] = \sum_{k=1}^K \lambda_k p_k$ is easier to evaluate than $\text{VaR}_{\alpha}[\mathcal{L}]$, which usually requires a Monte Carlo simulation.

We extend this simple uncertainty model to a more realistic one, where the defaults $X_k$ are no longer independent but follow a conditional independence scheme \cite{Rutkowski2014}. 
Given a realization $z$ of a latent random variable $\mathcal{Z}$, the Bernoulli random variables $X_k \mid \mathcal{Z}=z$ are assumed independent, but their default probabilities $p_k$ depend on $z$.
We follow \cite{Rutkowski2014} and assume that $\mathcal{Z}$ follows a standard normal distribution and that
\begin{eqnarray}
p_{k}(z) &=& F\left( \frac{F^{-1}(p_k^0) - \sqrt{\rho_k} z}{\sqrt{1 - \rho_k}} \right),
\end{eqnarray}
where $p_k^0$ denotes the default probability for $z=0$, $F$ is the cumulative distribution function (CDF) of the standard normal distribution, and $\rho_k \in [0, 1)$ determines the sensitivity of $X_k$ to $\mathcal{Z}$.
This scheme is similar to the one used for regulatory purposes in the Basel II (and following) Internal Ratings-Based (IRB) approach to credit risk \cite{BaselII, BaselIII}, and is called the \emph{Gaussian conditional independence model} \cite{Rutkowski2014}. 

In order to scale the model to a larger number of assets, one can aggregate subsets of similar assets into random variables $L_k \geq 0$ that take more than two values.
We briefly discuss this approach and the overall scaling of our algorithm to real world problems in Section \ref{sec:scaling}.

In the following sections, we show how the ECR for the presented model can be estimated on a gate-based quantum computer with QAE resulting in a quadratic speedup over classical Monte Carlo simulations.

\section{\label{sec:quantum_algorithm} Quantum Algorithm}

For the models introduced in Section \ref{sec:credit_risk_analysis}, the expected total loss $\mathbb{E}[\mathcal{L}]$ can be efficiently computed classically, see Appendix \ref{sec:expected_total_loss}.
Thus, we focus on quantum algorithms to estimate $\text{VaR}_{\alpha}[\mathcal{L}]$.
For more details on the estimation of expected values using QAE we refer to \cite{Woerner2019}.

To apply QAE, we map the problem of interest to a quantum operator $\mathcal{A}$ acting on $n+1$ qubits such that:
\begin{eqnarray}
\mathcal{A} \ket{0}_{n+1} &=& \sqrt{1 - a} \ket{\psi_0}_n \ket{0} + \sqrt{a} \ket{\psi_1}_n\ket{1},
\label{eq:a_operator}
\end{eqnarray}
where $a \in [0, 1]$. The probability to measure $\ket{1}$ in the last qubit, i.e., $a$, corresponds to the (normalized) property of interest.
From $\mathcal{A}$ we construct a quantum operator
\begin{eqnarray}
\mathcal{Q} &=& \mathcal{A} \mathcal{S}_0 \mathcal{A}^{\dagger} \mathcal{S}_{\psi_0},
\label{eq:q_operator}
\end{eqnarray}
where $S_0=\mathbb{I}-2\ket{0}_{n+1}\bra{0}_{n+1}$ and $\mathcal{S}_{\psi_0} = \mathbb{I}-2\ket{\psi_0}_n\ket{0}\bra{\psi_0}_n\bra{0}$.
Every application of $\mathcal{Q}$ corresponds to one \emph{quantum sample}.
QAE allows us to estimate $a$ with an estimation error that is bounded by
\begin{eqnarray}
\frac{2 \sqrt{a(1-a)}\pi}{M} + \frac{\pi^2}{M^2} &=& \mathcal{O}\left(\frac{1}{M}\right), \label{eq:ae_error}
\end{eqnarray}
where $M$ corresponds to the number of quantum samples \cite{Brassard2000, Woerner2019}.
QAE has a success probability of $81\%$, thus, by repeating only a few times and taking the median result the algorithm succeeds almost with certainty.
This leads to a quadratic speedup over classical Monte Carlo simulations, where the estimation error behaves as $\mathcal{O}(1/\sqrt{M})$, where $M$ now denotes the number of \emph{classical samples}.
A more detailed discussion of QAE can be found in Appendix \ref{sec:amplitude_estimation}.

To estimate VaR, we use QAE to efficiently evaluate the CDF of the total loss, i.e., we will construct $\mathcal{A}$ such that  $a = \mathbb{P}[\mathcal{L} \leq x]$ for a given $x \geq 0$, and apply a bisection search to find the smallest $x_{\alpha} \geq 0$ such that $\mathbb{P}[\mathcal{L} \leq x_{\alpha}] \geq \alpha$, which implies $x_{\alpha} = \text{VaR}_{\alpha}[\mathcal{L}]$ \cite{Woerner2019}.

\begin{figure}[hbtp]
\centering
\includegraphics[width=0.475\textwidth]{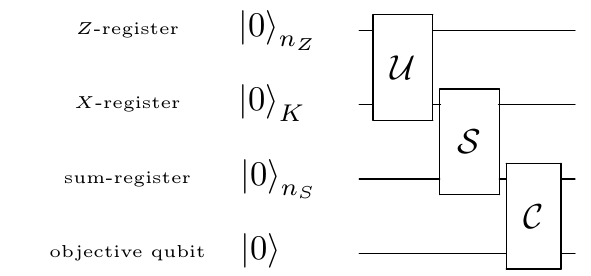}
\caption{\label{fig:high_level_cdf_circuit} High level circuit of the operator $\mathcal{A}$ used to evaluate the CDF of the total loss: the first qubit register with $n_Z$ qubits represents $\mathcal{Z}$, the second qubit register with $K$ qubits represents the $X_k$, the third qubit register with $n_S$ qubits represents the sum of the losses, i.e., the total loss, and the last qubit is flipped to $\ket{1}$ if the total loss is less than or equal to a given $x$.
The operators $\mathcal{U}$, $\mathcal{S}$, and $\mathcal{C}$ represent the loading of uncertainty, the summation of losses, and the comparison to a given $x$, respectively.}
\end{figure}

Mapping the CDF of the total loss to a quantum operator $\mathcal{A}$ requires three steps. Each step corresponds to a quantum operator.
First, $\mathcal{U}$ loads the uncertainty model. Second, $\mathcal{S}$ computes the total loss into a quantum register with $n_S$ qubits. Last, $\mathcal{C}$ flips a target qubit if the total loss is less than or equal to a given level $x$ which is used to search for $\text{VaR}_\alpha$.
Thus, we have $\mathcal{A} = \mathcal{CSU}$ and Fig.~\ref{fig:high_level_cdf_circuit} illustrates the corresponding circuit on a high level.

The estimation error given in Eq.~(\ref{eq:ae_error}) also depends on the exact result $a$.
In particular, if $a$ is close to $0$ or $1$ the constant in the error bound becomes very small.
When computing $\text{VaR}_\alpha$, we want to find the minimal threshold such that the estimated probability is larger than or equal to $\alpha$.
Thus, we can replace $a$ in Eq.~(\ref{eq:ae_error}) by $\alpha$ to get a better error bound.
When $\alpha = 99.9\%$ the error bound is approximately
\begin{eqnarray}
\frac{1}{5 M} + \frac{\pi^2}{M^2} \label{eq:ae_error_alpha}
\end{eqnarray}
which is independent of the other properties of the problem.
In other words, QAE is particularly good at estimating tail probabilities of distributions.

We now discuss the operators $\mathcal{U}$, $\mathcal{S}$, and $\mathcal{C}$ in more detail.
When the default events $\{X_1, ..., X_K\}$ are uncorrelated we can encode the $X_k$ of each asset in the state of a corresponding qubit by applying to qubit $k$ a $Y$-rotation $R_Y(\theta_p^k)$ \cite{Nielsen2010} with angle $\theta_p^k = 2\arcsin(\sqrt{p_k})$. Therefore the loading operator is
\begin{eqnarray}
\mathcal{U} &=& \bigotimes_{k=1}^K R_Y(\theta_p^k).
\end{eqnarray}
This prepares qubit $k$ in the state $\sqrt{1 - p_k}\ket{0} + \sqrt{p_k}\ket{1}$ for which the probability to measure $\ket{1}$ is $p_k$. 
The $\ket{1}$ state of qubit $k$ thus corresponds to a loss for asset $k$.

\begin{figure}[hbtp]
\centering
\includegraphics[width=0.45\textwidth]{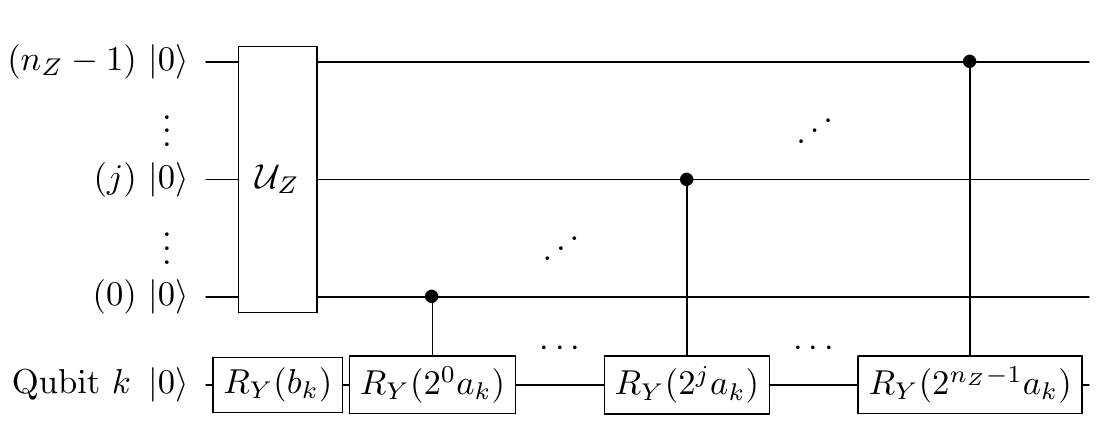}
\caption{\label{fig:affine_controlled_rotation} Affine dependency of $X_k$ on $\mathcal{Z}$: The qubit representing $X_k$ is prepared using $Y$-rotations controlled by the qubits representing $\mathcal{Z}$. Since the rotation angles are additive this construction rotates qubit $k$ by an angle $a_k z + b_k$.}
\end{figure}

To adjust $\mathcal{U}$ to include correlations between the default events, we add another register with $n_Z$ qubits to represent $\mathcal{Z}$.
The random variable $\mathcal{Z}$ follows a standard normal distribution. We use a truncated and discretized approximation with $2^{n_Z}$ values, where we consider an affine mapping $z_i = a_z i + b_z$ from $i \in \{0, ..., 2^{n_Z}-1\}$ to the desired range of values of $\mathcal{Z}$. 
Any discretized and truncated log-concave distribution, such as $\mathcal{Z}$, can be efficiently represented in a quantum register by an operator $\mathcal{U}_Z$ built from controlled rotations \cite{Grover2002}.
The qubit register representing $\mathcal{Z}$ is then used to control the rotation angles $\theta_p^k(z) = 2 \arcsin(\sqrt{p_{k}(z)})$ that prepare the qubits representing the $X_k$.
For simplicity, we use a first order approximation of $\theta_p^k(z)$ and include the affine mapping from $z$ (a value of the normal distribution) to $i$ (an integer represented by $n_Z$ qubits), i.e., $\theta_p^k(z_i) \approx a_k i + b_k$.
This affine dependency of the rotation angles $\theta_p^k$ with respect to $\mathcal{Z}$ can be constructed with a controlled rotation, see Fig.~\ref{fig:affine_controlled_rotation}.
Higher order approximations of $\theta_p^k(z)$ can be implemented using multi-controlled rotations. 
Furthermore, by using quantum arithmetic one could also compute $\theta_p^k(Y)$ directly \citep{Woerner2019}.

The ability to efficiently construct the uncertainty model is a crucial part in QAE-based algorithms, and if not handled carefully can diminish the potential quantum advantage.
The previous discussion shows that the Gaussian conditional independence model is particularly suitable for efficient loading in a quantum computer.
However, the depth of the circuit implementing $\mathcal{U}$, shown in Fig.~\ref{fig:affine_controlled_rotation}, scales as $\mathcal{O}(n_zK)$, i.e.~linear in the number of assets.
By adding $\mathcal{O}(K)$ ancilla qubits, the scaling of the circuit depth can be reduced to $\mathcal{O}(\log{K})$, which can lead to a potential speed-up.
The additional qubits provide the compute space to perform more operations in parallel.
Depending on the number of available qubits and the complexity of the rest of the algorithm, the number of ancillas can also be set to a smaller value to achieve an optimal overall performance.
The efficient implementation of $\mathcal{U}$ is discussed in detail in Sec.~\ref{sec:scaling}.

Next, we need to compute the resulting total loss for every realization of the $X_k$.
Therefore, we use a weighted sum operator
\begin{eqnarray}
&\mathcal{S}: & \ket{x_1, \cdots, x_K}_K \ket{0}_{n_S} \nonumber \\ 
&\mapsto & \ket{x_1, \cdots, x_K}_K \ket{ \lambda_1 x_1 + \cdots + \lambda_K x_K}_{n_S},
\end{eqnarray}
where $x_k \in \{0, 1\}$ denote the possible realizations of $X_k$.
We set $n_S = \lfloor\log_2(\lambda_1 + \cdots + \lambda_K)\rfloor + 1$ to represent in the second register all possible values of the sum of the losses given default $\lambda_k$, assumed to be integers.
An efficient implementation of $S$ is discussed in Sec.~\ref{sec:scaling}.

Last, we need an operator that compares a particular loss realization to a given $x$ and then flips a target qubit from $\ket{0}$ to $\ket{1}$ if the loss is less than or equal to $x$.
This operator is defined by
\begin{eqnarray}
\mathcal{C}: \ket{i}_{n_S}\ket{0} \mapsto 
\begin{cases}
\ket{i}_{n_S}\ket{1} & \text{if $i \leq x$}, \\
\ket{i}_{n_S}\ket{0} & \text{otherwise.}
\end{cases}
\end{eqnarray}
An efficient implementation of $\mathcal{C}$ is discussed in Sec.~\ref{sec:scaling}.

In the remainder of this paper we apply this algorithm to a small illustrative example using classical simulations of a quantum computer and we discuss the scaling to problems of realistic size.

\section{\label{sec:results} Results}


In this section, we analyze the performance of the quantum algorithm for an illustrative example with $K=2$ assets. 
The losses given default $\lambda_k$, the default probabilities $p_k^0$, and the sensitivities $\rho_k$ are given in Tab.~\ref{tab:illustrative_example}.
Within this section we set $n_Z=2$, and from the $\lambda_k$ it follows that $n_S = 2$.
Thus, $\mathcal{A}$ is operating on seven qubits that represent this problem on a quantum computer, including the objective qubit.

\begin{table}[htbp!]
\caption{\label{tab:illustrative_example} Problem parameters for the two-assets example.}
\begin{tabular}{cccc}
asset number & loss given default & default prob. & sensitivity \\
$k$ & $\lambda_k$ & $p_k^{0}$ & $\rho_k$ \\
\hline
1   &  1          & 0.15       & 0.1 \\
2   &  2          & 0.25       & 0.05
\end{tabular}
\end{table}

\begin{figure}[hbtp]
\centering
\includegraphics[width=0.475\textwidth]{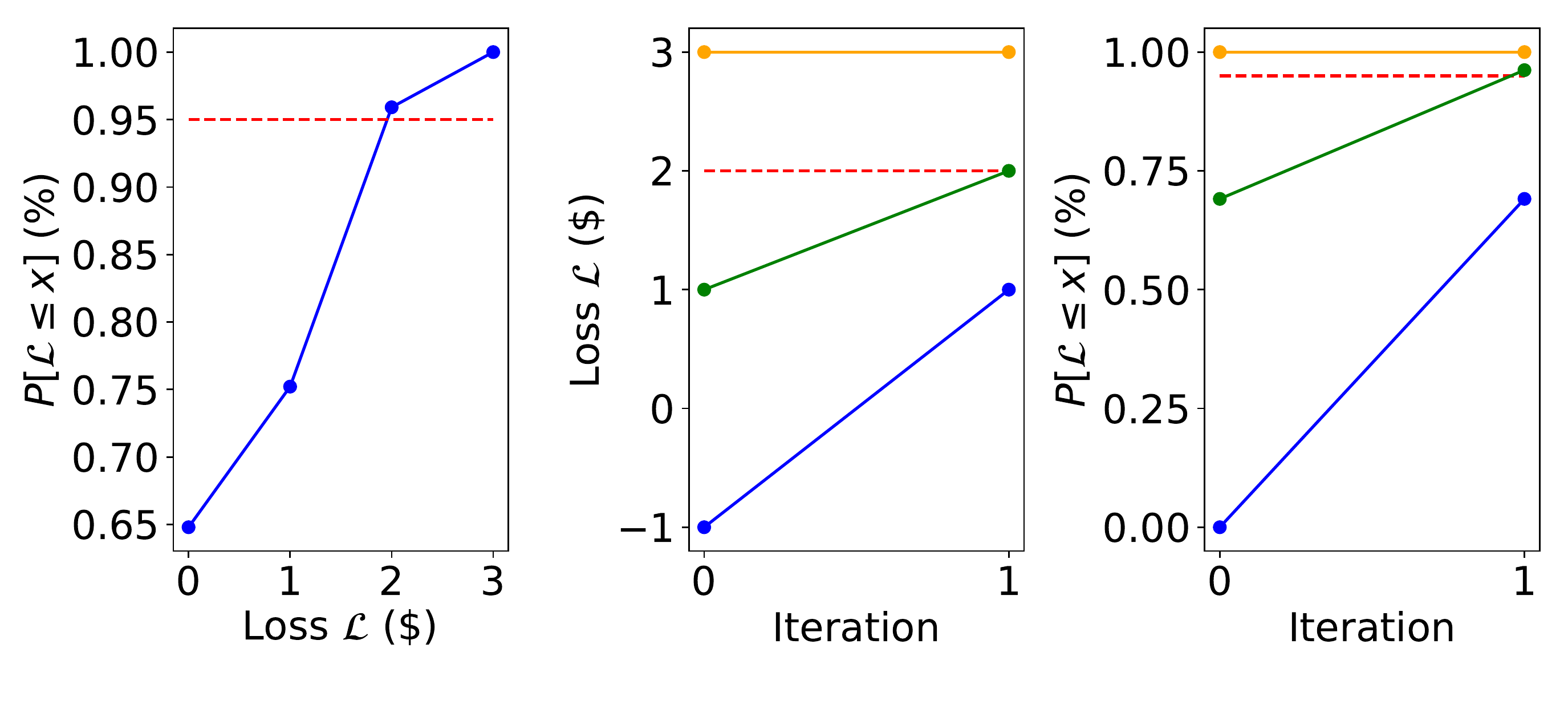}
\caption{\label{fig:var_bisection_search} 
Cumulative distribution function (left) of total loss $\mathcal{L}$ (blue) and target level of 95\% (red).
Bisection search to compute VaR (middle / right): Upper bound (orange), lower bound (blue), estimate (green), and exact value (red dashed line). Here, we set $\alpha = 95\%$ and $m = 4$.}
\end{figure}

To simulate our algorithm we input the circuit for $\mathcal{A}$ to the QAE sub-routine implemented in \emph{Qiskit}~\cite{Qiskit} and perform the bisection search using the result to find $x_\alpha$.
Since $n_S = 2$, the bisection search requires at most two steps, as shown in Fig.~\ref{fig:var_bisection_search}.
Note that QAE requires one additional ancilla  qubit to implement $\mathcal{Q}$ and we use four evaluation qubits giving us 16 quantum samples.
In total, this experiment requires 12 qubits that we simulate using classical computers.

\section{\label{sec:scaling} Scaling to real world problem}

We analyze the scaling of the quantum algorithm for problem sizes relevant to the finance industry. In particular, we analyze the circuit depth as a function of the number of assets $K$, to estimate the expected runtime on a fault-tolerant quantum computer~\cite{Shor1996, Kitaev2003, Fowler2012}.
We consider a gate decomposition into the Clifford + T gate set and mainly focus on the circuit depth in terms of T-gates, since they are the most expensive gates in a fault-tolerant quantum computer \cite{Bravyi2012}.
By using ancilla qubits, Toffoli gates can be constructed with a T-depth of one \cite{Selinger2013}, thus, we treat the two as equivalent in our runtime analysis.
Clifford gates, such as for instance CNOT-gates, are considered to be orders of magnitudes faster than T-gates and we mostly ignore them in the following \cite{Fowler2012, Fowler2018}.

Our algorithm mainly consists of $\mathcal{A}$, multiple applications of (controlled) $\mathcal{Q}$, and an inverse quantum Fourier transform (QFT) at the end.
The complexity of the inverse QFT scales at most quadratically with the number of evaluation qubits $m$, and is orders of magnitude smaller than the rest of the algorithm, since we assume $K \gg m$ and since the inverse QFT is only applied once.
Furthermore, the inverse QFT can even be approximated using $\mathcal{O}(n\log(n))$ T-gates \cite{Nam2018}, and, as discussed later in this section, it has been recently shown that Quantum Phase Estimation (QPE, includes the inverse QFT) can be omitted completely in QAE \cite{Suzuki2019}. 
Therefore, we ignore the contribution of the inverse QFT to the overall runtime.

Since the controlled powers of $\mathcal{Q}$ will dominate the runtime, we focus on the T/Toffoli-gates in $\mathcal{Q}$.
Eq.~(\ref{eq:q_operator}) implies that the controlled-$\mathcal{Q}$ operator in QAE requires controlling only the reflections $\mathcal{S}_0$ and $\mathcal{S}_{\psi_0}$. 
Indeed, $\mathcal{A}$ and $\mathcal{A}^{\dagger}$ are left uncontrolled and cancel each other when the control qubit of $\mathcal{Q}$ is in state $\ket{0}$, since in this case $\mathcal{S}_0$ is not applied.

We now argue that $\mathcal{S}_0$ and $\mathcal{S}_{\psi_0}$ do not dominate the runtime.
The reflection $\mathcal{S}_{\psi_0}$ can be implemented using an ancilla qubit and a phase kickback: an X-gate prepares the ancilla qubit in state $\ket{1}$, then the objective qubit of $\mathcal{A}$ is used to control a Z-gate targeting the ancilla qubit, a final X-gate uncomputes the ancilla qubit. This gate sequence transforms the objective qubit of $\mathcal{A}$ from $\alpha\ket{0}+\beta\ket{1}$ to $\alpha\ket{0}-\beta\ket{1}$ \cite{Nielsen2010} which is equivalent to the action of $\mathcal{S}_{\psi_0}$.
For a controlled application of $\mathcal{S}_{\psi_0}$ we replace the single-controlled Z-gate by a double-controlled Z-gate where the second control is an evaluation qubit.
Thus, $\mathcal{S}_{\psi_0}$ can be ignored in the overall runtime analysis as it can be implemented using a single Toffoli-gate (within the double-controlled Z-gate, exploiting that $Z = H X H$).

We implement $\mathcal{S}_0$ using the same construction as for $\mathcal{S}_{\psi_0}$ but with the single-controlled Z-gate replaced by a multi-controlled Z-gate that only acts if all qubits $\mathcal{A}$ operates on are in state $\ket{0}$.
However, if the sum-register is in state $\ket{0}_{n_S}$ then the $K$ qubits representing the $X_k$'s are also in state $\ket{0}_K$ and vice versa, since $\lambda_k > 0$ for all $k$.
Thus, instead of controlling the Z-gate with all state qubits, we only need to control it by the $n_Z$ qubits representing $\mathcal{Z}$, the $n_S$ qubits representing the total loss, and the objective qubit of $\mathcal{A}$.
Since multi-controlled gates can be implemented with logarithmic depth and a linear number of ancillas \cite{Maslov2015, Motzoi2017a}, we can also ignore the contribution of (controlled) $\mathcal{S}_0$ to the total runtime.

The previous discussion in this section implies that the multiple applications of $\mathcal{A}$ dominate the total runtime.
For $m$ evaluation qubits, $\mathcal{A}$ is called $n_S (2^{m+1} - 1)$ times: once for the initial state preparation, twice for each of the $2^m-1$ applications of $\mathcal{Q}$, and everything is repeated at most $n_S$ times for the bisection search to estimate VaR.

Since QAE is a probabilistic algorithm, we need to run it multiple times.
However, 25 repetitions are already sufficient to achieve a success probability of $99.75\%$ when using the median result \cite{Woerner2019}.
These are independent repetitions that could be parallelized on multiple separate quantum computers, thus, we do not include this additional overhead.

In the following, we analyze the circuit depth of $\mathcal{A}$.
How to efficiently implement the operators $\mathcal{U}$, $\mathcal{S}$, and $\mathcal{C}$ and the assumptions made, e.g., on approximation errors, is discussed in Appendices \ref{sec:uncertainty_loading}, \ref{sec:weighted_sum_operator}, and \ref{sec:fixed_value_comparator}, respectively.
The resulting circuit depths in terms of T/Toffoli-gates is stated in Table \ref{tab:scaling}.

\begin{table}[htbp!]
\begin{tabular}{cc}
operator & circuit depth (T/Toffoli-gates) \\
\hline
$\mathcal{U}$ & $26 + 28 n_Z$ \\
$\mathcal{S}$ & $\log_2(K) (\lfloor \log_2(n_S) \rfloor + \lfloor \log_2(n_S/3) \rfloor + 7)$\\
$\mathcal{C}$ & $2\lfloor \log_2(n_S - 1)\rfloor + 9$
\end{tabular}
\caption{Bounds on circuit depth of the operators $\mathcal{U}$, $\mathcal{S}$, $\mathcal{C}$ in terms of CNOT and Toffoli-gates, see Appendices \ref{sec:uncertainty_loading}, and \ref{sec:weighted_sum_operator}, \ref{sec:fixed_value_comparator}, for more details.}
\label{tab:scaling}
\end{table}

The total number of qubits will scale like $\mathcal{O}(K)$, since we represent every asset with a single qubit and the required ancillas also scale linearly in $K$.
We are mainly interested in an estimation of the overall runtime, and thus, we will not further elaborate on the exact number of required qubits.

In the remainder of this section, we consider $K = 2^{20}$, i.e., a portfolio of about one million assets, and assume $n_Z = 10$, and $n_S = 30$.
This implies that we discretize $\mathcal{Z}$ with $1,024$ different values, and that we assume the average of $\lambda_k$ is at most $1,024 =2^{n_S}/K$, otherwise $n_S$ would be too small to represent the maximal possible sum of losses.
Furthermore, we assume $m = 10$, which achieves an accuracy of $0.06\%$-points for $\alpha = 99.9\%$.

Inserting these numbers into the formulas in Table \ref{tab:scaling} leads to a T/Toffoli-depth for $\mathcal{A}$ of about $N_{\text{T}}^{\mathcal{A}} = 600$.
For the overall QAE, this implies a depth of
\begin{eqnarray} \label{eq:n_toffoli}
n_S (2^{m+1} - 1) N_{\text{T}}^{\mathcal{A}}.
\end{eqnarray}
which evaluates to a T/Toffoli-depth of approximately 37 million gates.

Up to now, we have not considered the impact of the limited connectivity of quantum processors, i.e., the fact that we need to introduce SWAP-gates to realize CNOT-gates or Toffoli-gates between qubits that are not physically connected.
It has been empirically shown in \cite{Woerner2019} for a related application that mapping comparable circuits to a realistic topology led to an increase in the number of CNOT-gates of about a factor of two.
Since the runtime is dominated by the time for the T-gates, doubling the number of CNOT gates in our circuit should not significantly affect the overall runtime. 
We therefore ignore the impact of limited connectivity.
Additionally, compiling the quantum circuits can be done in advance to produce a template circuit usable for a concrete problem.
Thus, the actual compilation time and circuit-optimization time is not added in our analysis.

We now assume that error-corrected T/Toffoli-gates can be executed in $10^{-4}$ seconds \cite{Fowler2018}. 
With this clock rate the 37 million gates obtained from Eq.~(\ref{eq:n_toffoli}) result in an estimated overall runtime of around one hour.
Removing the QPE from QAE not only allows to remove the inverse QFT but also reduces the overall circuit depth by a factor of two by allowing us to parallelize on two quantum devices \cite{Suzuki2019}.
This results in an estimated runtime of 30 minutes to estimate the VaR for a one-million-asset portfolio.

Classical simulations of large portfolios are a big computation problem which requires significant time and hardware resources \cite{Lan2010, Desmettre2016, Stockinger2018}.
To reduce classical simulation times, approximations are used and similar assets are aggregated in batches described by more complex random distributions.
The same methods can also be applied to our quantum algorithm and should be able to achieve similar improvements, potentially reducing the expected runtime of 30 minutes for one million assets to near real-time.
Furthermore, aggregating similar assets can also help to reduce the required number of qubits.

Unlike for classical algorithms, estimating the Conditional Value at Risk (CVaR, or Expected Shortfall) can be achieved without much additional overhead, since it is just one additional (slightly more expensive) application of QAE without the bisection search \cite{Woerner2019}.

\section{\label{sec:conclusion} Conclusion}

In this paper we developed and analyzed a quantum algorithm to estimate ECR with a quadratic speedup.
We have demonstrated the algorithm using a simulation and analyzed the scaling and expected runtime for realistic problem sizes under reasonable assumptions on future quantum computers.
Furthermore, we argued that our results also hold for more complex uncertainty models or other objectives, such as CVaR, without much additional overhead.
Although there is still a long way to go in terms of hardware development, this implies a huge potential for quantum computing in credit risk analysis.
Further research in algorithms can help to reduce the number of required qubits as well as the gate depth.

Within this paper, we made assumptions on the performance of future quantum hardware.
We tried to make our analysis as transparent as possible to allow adjustments of our results in case of new insights on future hardware or algorithmic components.
Until quantum computers of the required scale are available, a lot of research needs to take place also with focus on quantum algorithms, error correction, and circuit optimization.
Thus, it would not be surprising, if our assumption will turnout to be conservative, implying an even larger potential for the technology, than outlined in the present manuscript.

\begin{acknowledgments}
The authors want to thank Joan Francesc Vidal Villal\'on and Santiago Murillo Pavas from CaixaBank for the inspiring discussion on this important use cases, and James Wootton as well as Dmitri Maslov for their valuable insights on quantum error-correction and gate decomposition.
\end{acknowledgments}

\appendix

\section{Expected Total Loss} \label{sec:expected_total_loss}
In the following, we show how the total loss introduced in Section \ref{sec:credit_risk_analysis} can be efficiently computed classically.
For the first model in which the default events are independent the expected total loss is given by
\begin{eqnarray}
\mathbb{E}[\mathcal{L}] &=& \sum_{k=1}^{K} \lambda_k p_k.
\end{eqnarray}
This is due to the linearity of the expected value and the independence of the random variables $X_k$.

For the second model, we exploit the conditional independence assumption, which allows us to compute the expected loss as
\begin{eqnarray}
\mathbb{E}[\mathcal{L}] &=& \int_{z=-\infty}^{\infty} \sum_{k=1}^{K} \lambda_k p_k(z) f(z) dz,
\end{eqnarray}
where $f$ denotes the probability density function of the standard normal distribution.
This term can be efficiently approximated classically using numerical integration.

\section{Amplitude Estimation} \label{sec:amplitude_estimation}
\begin{figure}
\includegraphics[width=0.45\textwidth]{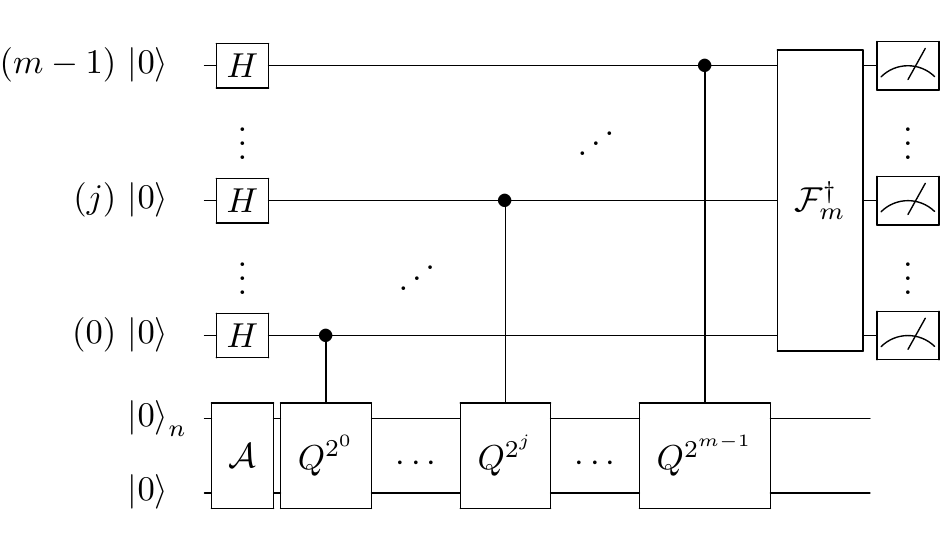}
\caption{The quantum circuit of amplitude estimation. \label{fig:ae}}
\end{figure}

The advantage for credit risk analysis comes from the quantum amplitude estimation (QAE) algorithm \cite{Brassard2000}, which provides a quadratic speed-up over classical Monte-Carlo simulations \cite{Abrams1999, Montanaro2017, Woerner2019, Stamatopoulos2019}.
Suppose a unitary operator $\mathcal{A}$ as defined in Eq.~(\ref{eq:a_operator}).
QAE allows the efficient estimation of $a$, i.e., the probability of measuring $\ket{1}$ in the last qubit. 
This estimation is obtained with an operator $Q$, given in Eq.~(\ref{eq:q_operator}), and Quantum Phase Estimation \citep{Kitaev1995}.
QAE requires $m$ additional evaluation qubits and $M = 2^m-1$ applications of $Q$.
The $m$ qubits, initialized to an equal superposition state by Hadamard gates, are used to control different powers of $Q$.
After applying the inverse Quantum Fourier Transform, their state is measured resulting in an integer $y \in \{0, ..., M-1\}$, which is classically mapped to the estimator $\tilde{a} = \sin^2(y\pi/M) \in [0, 1]$, see the circuit in Fig.~\ref{fig:ae}.
The estimator $\tilde{a}$ satisfies the error bound provided in Eq.~(\ref{eq:ae_error}) with probability of at least $8/\pi^2$.
This represents a quadratic speedup compared to the $\mathcal{O}\left(M^{-1/2}\right)$ convergence rate of classical Monte Carlo methods \citep{Glasserman2000}.

\section{Uncertainty Model} \label{sec:uncertainty_loading}
Every $X_k$-qubit needs to be prepared using an uncontrolled Y-rotation as well as $n_Z$ controlled Y-rotations.
On an error-corrected quantum computer Y-rotations can be realized with a T-depth of about $3\log_2(1/\epsilon)-4$ and controlled Y-rotations with a T-depth of about $3\log_2(1/\epsilon) - 2$, where $\epsilon > 0$ is the approximation error of the resulting unitary \cite{Amy2012, Kliuchnikov2012}.
Throughout this section we assume $\epsilon = 2^{-10} \approx 10^{-3}$, which implies a T-depth of 26 for uncontrolled Y-rotations and a T-depth of 28 for controlled Y-rotations.

A straight-forward implementation of $\mathcal{U}$ would require first $K$ uncontrolled Y-rotations followed by $n_Z K$ controlled Y-rotations --- from the $n_Z$ qubits representing $\mathcal{Z}$ to all $K$ qubits representing the $X_k$ --- with depth of $K$ controlled Y-rotations, since we can apply $n_Z$ rotations in parallel for $K \geq n_Z$.
In this analysis we ignore the preparation of $\mathcal{U}_Z$ as it can be done efficiently \cite{Grover2002} and does not depend on $K$, thus, has a negligible impact if $K \gg n_Z$.

A depth of $\mathcal{O}(K)$ is prohibitive for large portfolios due to the required T-gates.
To implement $\mathcal{U}$ more efficiently, we duplicate the $\mathcal{Z}$-qubits $(w-1)$-times, i.e., in total we have $w$ entangled copies of the $n_Z$ qubits representing $\mathcal{Z}$. This requires $n_Z (w-1)$ ancilla qubits and $2 n_Z w$ CNOT-gates, with a resulting CNOT-depth of $2 \log_2(w)$, since every entangled copy can be reused to prepare more copies. The factor $2$ appears since we should uncompute this preparation at the end.
Having $w$ copies of $\mathcal{Z}$ allows us to parallelize the preparation of the $X_k$-qubits, achieving a depth of $n_Z K/w$ controlled Y-rotations.
To minimize the T-depth, we set $w = K$, i.e., we add $n_Z(K-1)$ entangled copies with a CNOT-depth of $\log_2(K)$, leading to a Y-rotation-depth of $n_Z$, independent of $K$.
Combining this with the T-depth for Y-rotations leads to total T-depth for $\mathcal{U}$ of $26 + 28 n_Z$.

\section{Weighted Sum Operator} \label{sec:weighted_sum_operator}
Next, we analyze the implementation of $\mathcal{S}$.
Again, we can significantly reduce the circuit depth using additional ancilla qubits.
We apply a divide and conquer approach and first sum up pairs of assets, then pairs of the resulting sums and so on until we computed the total sum.

This implies that we start with a weighted-sum operator as outlined in Sec.~\ref{sec:quantum_algorithm} and also introduced and discussed in detail in \cite{Stamatopoulos2019}, and then continue with adder circuits \cite{Cuccaro2004} to iteratively combine the intermediate results.

For simplicity, we consider average values in the following analysis of the circuit depth.
Since we have $K$ assets and assume the total loss can always be represented with $n_S$ qubits, the loss per asset can be represented using on average at most $\log_2(2^{n_S} / K) = n_S - \log_2(K)$ qubits.
On average, the intermediate values at most double from one iteration to the next, i.e. we need to add at most one qubit per iteration to each intermediate result and after $\log_2(K)$ iterations we have computed the total loss.

Within every iteration, we assume that the individual intermediate results can be computed in parallel.
For iteration $i$, $i = 0, \ldots, \log_2(K)-1$, we assume the values are represented each by $n_i = (n_S - \log_2(K) + i)$ qubits. An adder circuit on $n$ qubits can be realized with a Toffoli-depth of  
\begin{eqnarray}
\lfloor \log_2(n) \rfloor + \lfloor \log_2(n/3) \rfloor + 7,
\end{eqnarray}
using a linear number of ancilla qubits \cite{Draper2004}.
Thus, we can bound the overall Toffoli-depth by 
\begin{eqnarray}
\log_2(K) (\lfloor \log_2(n_S) \rfloor + \lfloor \log_2(n_S/3) \rfloor + 7).
\end{eqnarray}

\section{Fixed Value Comparator} \label{sec:fixed_value_comparator}

A fixed value comparator, i.e., a comparator that takes a fixed value to compare to as a classical input, can be based on adder circuits.
The result reported in Table \ref{tab:scaling}, i.e., a Toffoli-depth of $2\log_2(n_S - 1) + 9$, is taken from \cite{Draper2004} where a construction of adders and comparators is introduced and analyzed in detail.
To achieve the logarithmic scaling, a linear number of ancilla qubits needs to be added.

\bibliography{quantum_credit_risk_analysis}

\end{document}